\documentclass[11pt, a4paper, oneside]{article}
\usepackage[T1]{fontenc}
\usepackage[utf8]{inputenc}
\usepackage{authblk}
\usepackage{bbm}
\usepackage{amssymb,amsmath}
\usepackage{tikz}
\usepackage{todonotes}
\usepackage{subfig}
\usepackage{cite, hyperref}
\usepackage[margin=1in]{geometry}

\usetikzlibrary{arrows, decorations.markings}
\usetikzlibrary{decorations.pathreplacing}
\usepackage{algpseudocode,algorithm,algorithmicx}
\usepackage{comment}

\usepackage{amsthm}
\theoremstyle{definition}

\theoremstyle{theorem}

\theoremstyle{remark}

\theoremstyle{example}

\newcommand{\R}{\mathbb{R}}

\newcommand{\ie}{\textit{i.e.}}

\newcommand{\ii}{\mathbf{i}}
\newcommand{\X}{\mathbf{X}}

\newcommand{\dd}{\mathrm{d}}
\newcommand{\uu}{\mathbf{u}}

\providecommand{\keywords}[1]{\noindent \textit{Keywords:} #1}

\date{\today}

\title{Introducing higher order correlations to marginals' subset of 
	multivariate data by means of 
	Archimedean copulas}
\author[1]{Krzysztof Domino\thanks{kdomino@iitis.pl}}
\author[1,2]{Adam Glos}
\affil[1]{Institute of Theoretical and Applied Informatics,\protect\\
	Polish Academy of Sciences,\protect\\
	Ba{\l}tycka 5, 44-100 Gliwice, Poland}
\affil[2]{Institute of Informatics, Silesian University of Technology,\protect\\
	ul. Akademicka 16, 44-100 Gliwice, Poland}
\begin{document}

\maketitle
\begin{abstract}
In this paper, we present the algorithm that alters the subset of marginals of
multivariate standard distributed data into such modelled by an Archimedean
copula. Proposed algorithm leaves a correlation matrix almost unchanged, but
introduces a higher order correlation into a subset of marginals. Our data
transformation algorithm can be used to analyse whether particular machine
learning algorithm, especially a dimensionality reduction one, utilises higher
order correlations or not. We present an exemplary application on two features
selection algorithms, mention that features selection is one of the approaches
to dimensionality reduction. To measure higher order correlation, we use
multivariate higher order cumulants, hence to utilises higher order correlations
be to use the Joint Skewness Band Selection (JSBS) algorithm that uses
third-order multivariate cumulant. We show the robust performance of the JSBS in
contrary to the poor performance of the Maximum Ellipsoid Volume (MEV) algorithm
that does not utilise such higher order correlations. With this result, we
confirm the potential application of our data generation algorithm to analyse a
performance of various dimensionality reduction algorithms.
\end{abstract}


\section{Introduction}

While analysing real-life multivariate data such as financial data, e-commerce 
data, biomedical data, audio signals or high resolution 
images~\cite{fan2011sparse, ando2017clustering, dai2015feature, 
li2017clustering, cordes2015feature}, 
we have many features that carry valuable information.
However, if a number of features is large, the computational cost of 
proceeding such data is
high. To transfer such data to a smaller set of features, in such a way
that most of the meaningful information is preserved, we can use a
dimensionality reduction scheme \cite{roweis2000nonlinear}. Many dimensionality
reduction schemes asses the importance of features by analysing their
correlation or covariance. Hence, assuming that data follow a multivariate
Gaussian distribution \cite{duda2012pattern}. Such methods may be ineffective if
dealing with information hidden in higher order correlations. As an example, one
can consider the Principal Component Analysis (PCA) \cite{jolliffe2002principal}
that converts data into a subset of independent features via linear
transformations determined by the eigenvectors of a covariance matrix. The PCA
works well on data for which the second order correlations dominate. However,
the real data may possess higher order dependencies \cite{jondeau2017moment,
arismendi2014monte, becker2014eeg, geng2011research}, which in turn may     
change the optimal result.

While it is usually easy to determine how higher correlations are considered for
statistics based algorithm, in more advanced machine-learning algorithms it is 
not obvious to what extend higher order correlations influence the output. As a 
straight example 
consider a Kernel enhanced PCA \cite{scholkopf1998nonlinear, 
hoffmann2007kernel} or Kernel enhanced discriminant analysis 
\cite{baudat2000generalized}, where the method's utility depends on the 
particular choice of the Kernel function given a particular data set. 
For further discussion of non linear dimensionality reduction algorithms see 
also \cite{lee2007nonlinear}. 
There is as well a neural network approach to dimensionality reduction 
\cite{sakurada2014anomaly} where a
dimensionality reduction for an outlier detection was performed by means of a
neural networks on both artificially generated and real data. Another neural 
networks example that uses auto-encoders is discussed in 
\cite{wang2014generalized, hinton2006reducing}.
The properly configured neural network (using auto-encoders) detects subtle
anomalies successfully, but the PCA fails. This suggest the sensitivity of such 
neural networks on higher order correlations.

The main contributions of the presented work is the algorithm transforming
multivariate Gaussian distributed data into non-Gaussian distributed one. We 
argue that such 
algorithm may be applicable 
to analyse various dimensionality reduction algorithms. More precisely, we 
present a method of transforming Gaussian distributed data into such with 
higher order correlations inside a chosen subset of marginals, but with a 
covariance (second order correlations) similar to those of original data. This 
is done by means of various Archimedean copulas \cite{mcneil2009multivariate}, 
hence in addition all univariate marginal distributions are unchanged.
In order to show that our method works properly, we have tested it on two 
distinct features selection algorithms: the MEV 
(Maximum Ellipsoid Volume) \cite{sheffield1985selecting}, that select features 
(a subset of marginals) on a basis of second order correlations, and the JSBS 
(Joint Skewness Band
Selection)~\cite{geng2015joint}, that select features on a basis of third 
order correlations. To measure this correlations we use second and third order 
multivariate cumulants. We show that in contrary to MEV the JSBS can detect a 
subset of marginals that exhibit higher order correlations. The MEV is 
ineffective since the covariance matrix is almost unchanged by a data 
transformation.

To 
provide the tool for machine-learning scientists, the presented
algorithm is implemented in the 
\texttt{Julia} programming language \cite{bezanson2014julia} and is available 
on a GitHub repository \cite{cop}. The \texttt{Julia} is high level programming 
language, suitable for scientific computations, because it is open source and 
the code can be analysed and reviewed by scientist. Apart from this,
linear operations and random sampling operations implemented in the 
\texttt{Julia} takes significantly less of the processor time than similar 
operations implemented in other well known programming languages, see 
\cite{bezanson2014julia}. Finally \texttt{Julia} is easily accessible from 
\texttt{Pyhon} containing a large collection of machine-learning tools.

The paper is organized as follows. In Sec.~\ref{sec:preliminaries} we provides
basic facts concerning multivariate distribution and feature discrimination
algorithms. In Sec.~\ref{sec:algorithm} we present and analyse our algorithm
using MEV and JSBS algorithms. In Sec.~\ref{sec:conclusions} we sum up our
results and discuss its applications and extensions.

\section{Preliminaries} \label{sec:preliminaries}

To measure higher order correlations between features we use higher order 
multivariate cumulants. A multivariate cumulant is a multivariate 
extension of corresponding univariate cumulant \cite{mccullagh2009cumulants}. 
As such a multivariate cumulant of order $d$ can be represented in a form of a 
super-symmetric tensor $M^d\in 
\R^{[n,d]}$~\cite{schatz2014exploiting, domino2017tensorsnet}. Such 
cumulant's tensor is the $d$-dimensional array indexed by $\ii = (i_1, 
\ldots, i_d)$. The super-symmetry means here, that each 
permutation within $\ii$ gives the index that refers to the same value of the 
tensor as $\ii$. Note that 
the first cumulant is an expectation and the second is 
a covariance matrix, that both fully describe Gaussian multivariate 
distribution. 
Importantly for Gaussian multivariate distribution cumulants of order higher 
than $2$, 
called higher order cumulants, are zero 
\cite{kendall1946advanced,lukacs1970characteristics}. Concluding we have zero 
higher order correlations there.

Non-zero higher order correlations can be introduced by 
means of copulas. A copula $C(\uu): [0,1]^n \rightarrow [0,1]$ is a join 
multivariate cumulative 
distribution function with uniform marginal on a $[0,1]$ segment 
\cite{nelsen1999introduction}. A sub-copula $C_{\mathbf r}$ at 
index $\mathbf r = (r_1,\dots,r_k)\subset (1,\dots,n)$ is a
joint multivariate cumulative distribution of a $\mathbf r$ subset of 
marginals. There is an important family of copulas called Archimedean 
copulas, recently used to model various types of real life data such as: 
financial data
\cite{cherubini2004copula, embrechts2001modelling, naifar2011modelling,
	domino2014use}, hydrological data \cite{zhang2012application, 
	tsakiris2015flood}, signals 
\cite{zeng2014copulas}, wireless communication data
\cite{peters2014communications} or biomedical data 
\cite{silva2014statistically}. The Archimedean copula is introduced by 
a copula generator function $\psi$. 

\begin{figure}[t]\centering
	\subfloat[Clayton]{\includegraphics[]{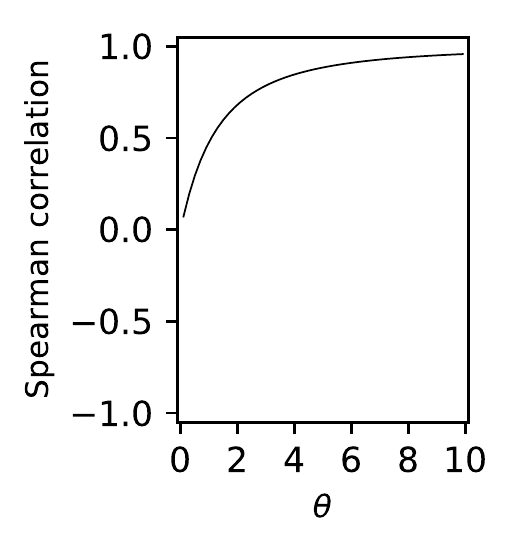}}
	\subfloat[Gumbel]{\includegraphics[]{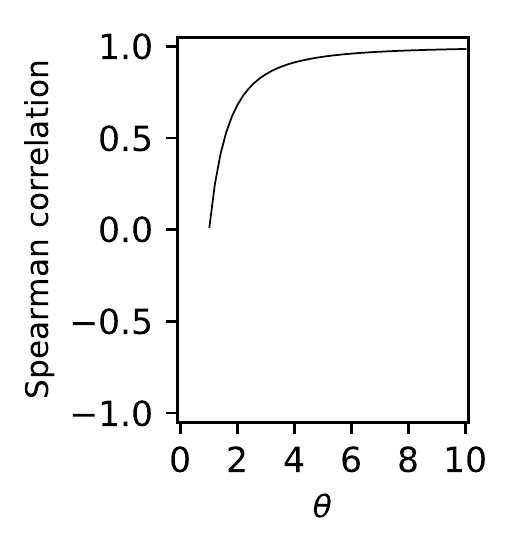}}
	\subfloat[AMH]{\includegraphics[]{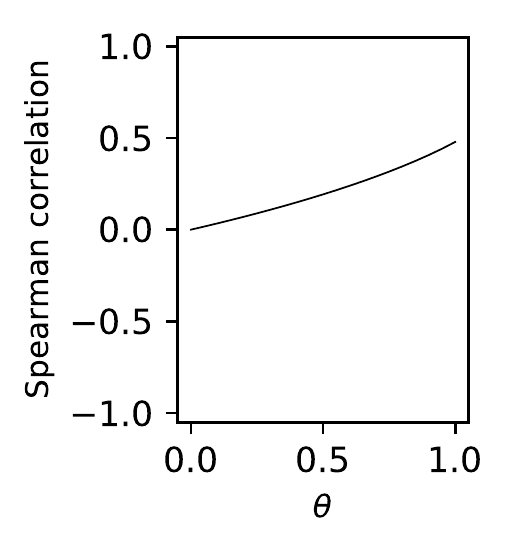}}
\caption{Relation between copula's parameter $\theta$ and Spearman's 
correlation given by Eq.~\eqref{eq::cor} for various Archimedean copulas.}	
\label{f::cors}
\end{figure}

Let 
$\psi_\theta:[0,\infty)\to[0,1]$ be
continuous function, parametrised by $\theta$, such that $\psi_\theta(0) = 1$ 
and $\psi_\theta(\infty) =
0$. Furthermore let $\psi_\theta$ be a strictly decreasing function on $[0,
\inf\psi_\theta^{-1}(\{0\})]$, where $\psi_\theta^{-1}$ is its pseudoinverse
fulfilling $\psi^{-1}(0) = \inf \{u: \psi(u) = 0\}$. Finally let
$\psi_{\theta}$ be the $n$-monotone in $[0, \infty)$ according to
Definition~$2.3$ in \cite{mcneil2009multivariate}. Having introduced the 
generator function, the Archimedean copula
takes the following form
\begin{equation}\label{eq::archcop}
[0,1] \ni C(\uu) = \psi\left (\sum_{i=1}^{n} \psi^{-1}(u_i) \right ),
\end{equation}
where $\uu \in [0,1]^n$. Well known 
examples of Archimedean copulas are Gumbel, Clayton and
Ali-Mikhail-Haq (AMH) \cite{kumar2010probability} ones. All of them are 
parametrized by a single real-value parameter $\theta$. A properties of those 
copulas are presented in Table~\ref{table:archimedean-examples}.
The Spearman's correlation between marginal variables 
$\mathcal{X}_{i_1},\mathcal{X}_{i_2}$ 
modelled by a copula $C$ takes the form~\cite{schweizer1981nonparametric}
\begin{equation}\label{eq::cor}
\rho(\mathcal{X}_{i_1},\mathcal{X}_{i_2}) =  12\int_0^1\int_0^1 
C_{(i_1,i_2)}(u_1,u_2)  \dd u_1 \dd u_2 -3,
\end{equation}
where $C_{(i_1, i_2)}$ is a sub-copula for marginal variables 
$\mathcal{X}_{i_1}$ and $\mathcal{X}_{i_2}$. 
For copulas presented in Tab.~\ref{table:archimedean-examples}
the Spearman's correlation depends monotonically on the $\theta$ parameter, and 
hence uniquely determines it, see Fig.~\ref{f::cors}. However for Archimedean 
copulas such bivariate Spearmann's correlation do not carry all information 
about dependency between marginals.

\begin{figure}[t!]
	\centering \subfloat[Clayton
	copula.\label{fig::clc3}]{\includegraphics{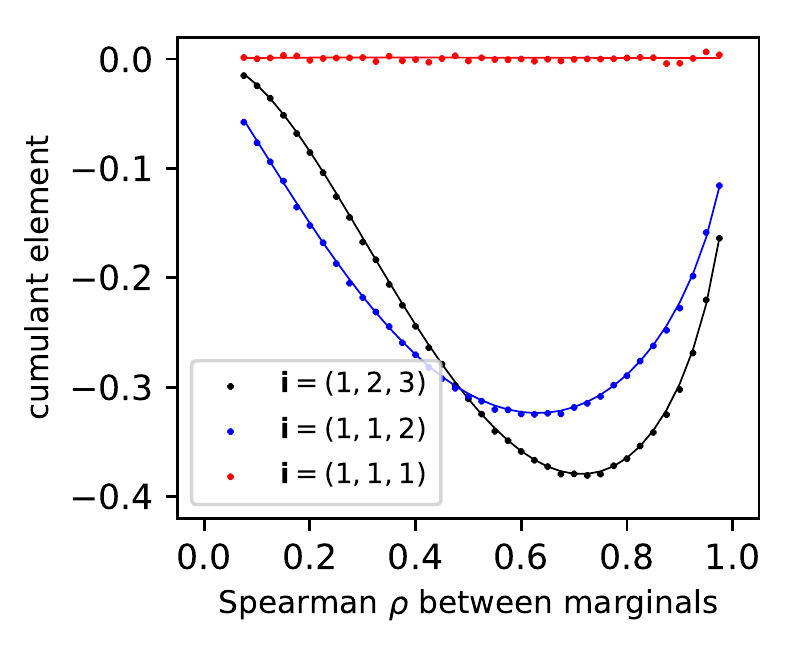}}
	\subfloat[Gumbel
	copula.\label{fig::guc3}]{\includegraphics{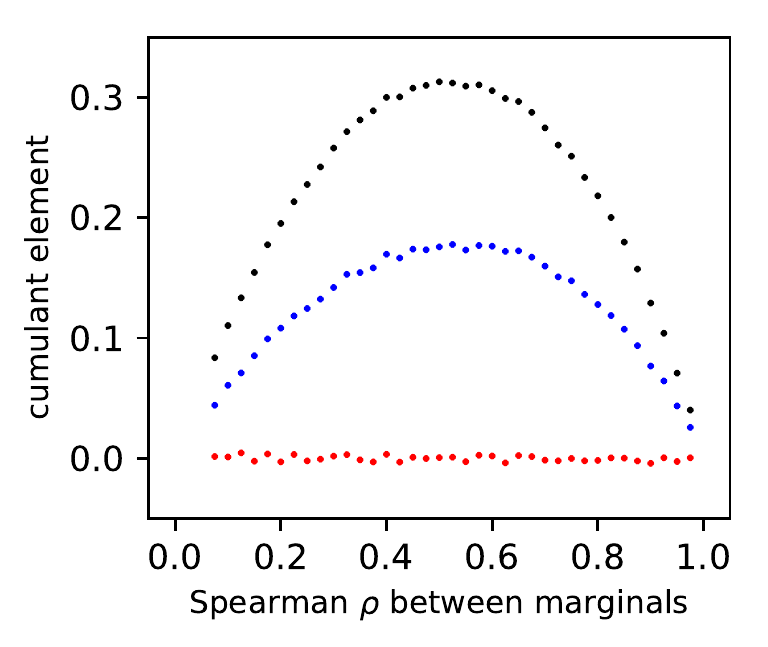}} \\
	\subfloat[AMH
	copula.\label{fig::amhc3}]{\includegraphics{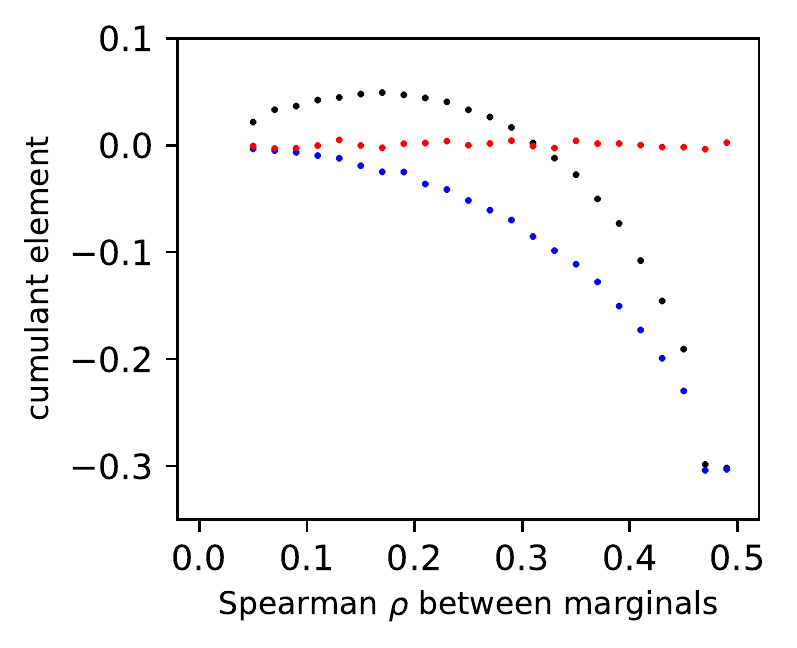}}
	\caption{Distinguishable elements of $3$\textsuperscript{th} cumulant 
	tensor for
		Archimedean copulas with Gaussian standard univariate marginals, 
		computed 
		for $t=10^6$ samples. Due
		to symmetry properties of the Archimedean copulas only $3$ elements are 
		distinguishable: the
		super-diagonal element at index $\ii = (1,1,1)$, the partial-diagonal 
		element at index $\ii = (1,1,2)$ and the off-diagonal element at index 
		$\ii 
		= (1,2,3)$. In a case of the Clayton
		copula we present theoretical outcomes \cite{de2012moments} that are 
		consistent
		with simulation ones, errors of simulation outcomes are
		negligible. The super-diagonal element values are zero due to 
		univariate 
		Gaussian marginals. Apart from this other cumulants' elements are 
		non-zero.}
	\label{fig::third-cumulants} \centering
\end{figure}

In contrary to multivariate Gaussian distribution, 
higher order cumulants for Archimedean copula are not necessarily zero 
tensors, see Fig.~\ref{fig::third-cumulants}. This fact implies higher order 
dependence between marginals. Due to its symmetry, an Archimedean
copula with identical univariate marginal distributions produces only three 
distinct elements of the $3$\textsuperscript{rd} cumulant's tensor. Those are 
the super-diagonal element, the partial-diagonal one and the off-diagonal one. 
The covariance matrix of such copula model that have only 
two distinct elements: the diagonal one and the off-diagonal one. 

\begin{table}[t]
	\caption{Definition of Gumbel, Clayton and Ali-Mikhail-Haq copulas, and 
		a possible correlation values. Note that Spearman's correlation $\rho$ 
		does not depend on choice of univariate marginal 
		distributions.}
	\label{table:archimedean-examples}
\begin{tabular}{llll}
	\hline\noalign{\smallskip}
	Copula name		& $\theta$ values	& $\psi_\theta(t)$ & $\rho$ values \\
	\noalign{\smallskip}\hline\noalign{\smallskip}
	Gumbel  		& $[1,\infty)$		&  
$\exp(-t^{\frac{1}{\theta}})$  & $[0,1)$ \\
Clayton  		& $(0,\infty)$		& $(1+ 
t)^{\frac{-1}{\theta}}$ & $(0,1)$ \\
 AMH 				& $(0,1)$			& 
$\frac{1-\theta}{\exp(t)-\theta}$ & $(0,1/2)$ \\
\noalign{\smallskip}\hline
\end{tabular}
\end{table}

For sampling Archimedean copulas we use modified Marshall-Olkin Algorithm. The
original algorithm is presented in Alg.~\ref{alg:arch_sampler}. Note that the
algorithm requires sample of inverse Laplace–Stieltjes transform
$L_{\psi_\theta}^{-1}$ of the Archimedean copula generator $\psi_\theta$.

\begin{algorithm}[t]
	\caption{Marshall-Olkin algorithm~\cite{marshall1988families, 
	hofert2008sampling} sampling an Archimedean copula.}
	\label{alg:arch_sampler}
	\begin{algorithmic}[1]	
		\State \textbf{Input}: $\psi_\theta$ -- generating function of the 
		Archimedean  copula, $\mathbf{x}$ -- $k$ samples of $U(0,1)$ 
		distribution, 
		$y$ -- sample of $U(0,1)$ distribution.
		\State \textbf{Output:} sample of Archimedean copula. 
		\Function{arch\_sampler}{$\psi_\theta$, $\mathbf{x}$, $v$}
		\State $v = \textsc{quantile}_{L^{-1}(\psi_\theta)}(y)$
		\State $k = \text{length}(\mathbf x)$
		\For{$i=1,\ldots,k$}
			\State $u_{i} \leftarrow \psi_\theta\left ( \frac{-\log 
			(x_{i})}{v}\right ) $
		\EndFor
		
		\State\Return $(u_1,\dots,u_k)$
		
		\EndFunction 
	\end{algorithmic}
\end{algorithm}

\subsection{MEV and JSBS algorithms}

Having introduced multivariate cumulants we can discuss now cumulants based 
features selection algorithms.
Let $\R^{t \times n} \ni \X = (X_1, \ldots, X_n)= (\mathbf{x}_1, \ldots,
\mathbf{x}_{t})^{\intercal}$ be the $n$ dimensional sample from the random
vector $\mathcal{X} = (\mathcal{X}_1,\dots,\mathcal{X}_n)$. Here $X_i =
(x_{1,i}, \ldots x_{t,i})^{\intercal} \in \R^t$ is a vector of $t$ realisations
of $i$\textsuperscript{th} feature (marginal) while $\mathbf{x}_{l} = (x_{l,i},
\ldots x_{l,n}) \in \R^n$ a single realisation of $\mathcal{X}$. The feature
selection algorithm chooses a subset
$\{\mathcal{X}_{i_1},\dots,\mathcal{X}_{i_k}\}$, which should provide as much
information as possible comparing to original $\mathcal{X}$. Alternatively the
feature selection problem can be understood as providing new order of
marginals $(\mathcal{X}_{j_1},\dots \mathcal{X}_{j_n})$, which represents their
importance. Then for fixed $k$  subset
$\{\mathcal{X}_{j_1},\mathcal{X}_{j_2},\dots,\mathcal{X}_{j_k}\}$ are a
representative collection of the original $\mathcal{X}$.

MEV (Maximum Ellipsoid Volume) \cite{sheffield1985selecting} is a feature
selection algorithm, that iteratively removes the least informative
variable. The choice bases on the maximisation of a hyper-ellipsoid volume in a
eigenvector space of a covariance matrix of reminding marginals. In details for
each marginal variable $\mathcal{X}_i$ the algorithm computes a determinant of
subcovariance matrix which is constructed from the original one by removing
$i$-th column and row. The variable, which provides smallest determinant is
considered to be the least informative at this point. Then the MEV algorithm 
recursively 
searches
for consecutive variable in remaining collection. This procedure provides the 
information order of marginal variables. However, as described above, 
the MEV
algorithm is based on the covariance matrix being a second cumulant of
multivariate data. Hence if for some subset of marginals cumulant of higher
order is non-zero, it may be ignored by the MEV algorithm.

The JSBS \cite{geng2015joint} algorithm is a natural extension of MEV algorithm,
which analyse the third-order multivariate cumulant. Since such cumulant can be 
represented as the $3$-mode
tensor, for which determinant is not well defined, in \cite{geng2015joint}
authors have optimized the following target function:
\begin{equation}
JS(\mathbf X) = \sqrt{\frac{\det\left 
[\mathbf{M^3_{(1)}}(\mathbf{M^3_{(1)}})^\top\right 
]}{\det^3 (\mathbf{M}^2) }},
\end{equation}
where $\mathbf{M_{(1)}^3}$ is an unfolded third cumulant $M^3$ in mode $1$ and 
$\mathbf{M}^2$ is 
a covariance matrix. As cumulants
are super-symmetric \cite{schatz2014exploiting, domino2017tensorsnet}, it is not
important in which mode they are unfolded. The $JS$ (Joint Skewness) can 
be interpreted as
product of singular values taken from HOSVD of the third cumulant's tensor 
divided by the product of the covariance matrix eigenvalues raised to power $3$.

\section{The algorithm} \label{sec:algorithm}
In this section we propose and analyse algorithm, which transforms part of the 
normally distributed data. Our goal is to replace part of the originally Gaussian data by samples
distributed according to various copulas. Our algorithm allows replacing
arbitrary subset of marginal variables of size $k$ by chosen Archimedean
copula. In particular we focused on copulas presented in
Sec.~\ref{sec:preliminaries}.  We use proposed algorithm to show difference 
between detection using the MEV and the JSBS algorithm. 

\subsection{Data malformation algorithm}\label{sub::datamalf}

Suppose we have $\X \in \R^{t \times n}$ that are $t$ samples from Gaussian 
$n$-variate 
distribution
and we want to replace all realisations of $i_1, \ldots, i_k$ marginals,
\ie~$X_{i_1}, \ldots, X_{i_k}$ by $X'_{i_1}, \ldots, X'_{i_k}$ modelled by
Archimedean $k$-variate copula, but leave unchanged other marginals \ie~those
indexed by $\{1,\ldots,n\} \setminus \{i_1, \ldots, i_k\}$. If we denote new
data by $\X'$, ideally we would expect $\text{cor}(\X) = \text{cor}(\X')$ to
make a transformation hard to detect by methods using the second order
correlations only.

Our algorithm, presented in Alg.~\ref{alg::malf_gen}, takes several steps: first
data are standardized, so all marginal variables have zero mean and variance 
$1$.
Then based on these we produce new $t$ samples of $(k+1)$-variate random vector 
with independent marginals, all distributed according to $U(0,1)$~\ie~uniform 
distribution on $[0,1]$. Those new core samples are 
generated in an information
preserving way using Alg.~\ref{alg::core} or naively for reference. Then the
parametrization $\theta$ of Archimedean copula is derived. New samples are
produced by means of Alg.~\ref{alg:arch_sampler} given core samples and the 
$\theta$ parameter. Finally original univariate 
distributions (Gaussian in our case) are recovered.


Let us first focus on the parameter $\theta$ derivation. While several
approaches can be considered here, our method bases on the fact that correlation
matrix of the Archimedean copula is constant outside the diagonal. We calculate
the mean value of an upper-triangle of the Spearman's correlation matrix of data
$[X_{i_1},X_{i_2},\dots,X_{i_k}]$, and then recover the $\theta$ value thanks to
Eq.~(\ref{eq::cor}).

Now let us consider the second part of the algorithm, which is presented in
Alg.~\ref{alg::core}. Observe first, that Alg.~\ref{alg:arch_sampler} converts 
$k+1$ samples
of $U(0,1)$ distribution into a sample of $k$-variate Archimedean copula. The
naive approach would be to generate independently elements of $\mathbf 
X^{\rm core} \in [0,1]^{t \times (k+1)}$ from $U(0,1)$ distribution, and
transform them using Alg.~\ref{alg:arch_sampler} to $t$ samples of $k$-variate 
Archimedean copula. Finally one transforms univariate 
marginals by quantile functions of original univariate frequency distributions 
(Gaussian in our case) to $X'_{i_1} , \ldots, X_{i_k}'$. Such naive
approach preserves univariate marginal distributions due to the copula 
approach, and roughly 
preserves a
correlation inside a subset $i_1, \ldots, i_k$ if copula parameter is chosen 
properly, but results in almost no
correlation between changed and not-changed marginals. This implies
$\text{cor}(\X) \not\approx \text{cor}(\X')$, which would made a detection easy
for methods based on the second order correlation (the covariance matrix).

To overcome this problem we need to collect information about the general
correlation between subsets of $k$ changed and $n-k$ non changed marginals and
input it into Alg.~\ref{alg:arch_sampler}. To store such information we use the
$(k+1)^{\rm th}$ column of $\X^{\rm core}$ matrix~\ie~the vector 
$X^{\text{core}}_{k+1}$ and substitute its elements for $y$ in 
Alg.~\ref{alg:arch_sampler}. In this algorithm we use a function
\begin{equation}\label{eq::fv}
	f(x_i, v) = \psi_{\theta}\left(\frac{-\log(x_i)}{v} \right)
\end{equation}
to compute a sample of $i$\textsuperscript{th} marginal of the Archimedean 
copula. By the Archimedean copula generator definition, such function
is strictly increasing in $v$ for an arbitrary constant $x_i \in (0,1)$. 
Furthermore, $v$ is a sample
of inverse Laplace–Stieltjes transform of the copula generator
$L^{-1}_{\psi_{\theta}}$, which is the CDF function for the range of $\theta$ 
considered in this paper, see Tab.~\ref{table:archimedean-examples} and 
\cite{hofert2008sampling}.
Its inverse (a quantile function) $\textsc{quantile}_{L^{-1}(\psi_\theta)}$ is 
strictly
increasing if $L_{\psi_{\theta}}^{-1}$ is continuous and non-decreasing if
$L_{\psi_{\theta}}^{-1}$ is discrete. Hence the composition
$f(\textsc{quantile}_{L^{-1}(\psi_\theta)}(y))$ is strictly
increasing for continuous $L_{\psi_{\theta}}^{-1}$, and non-decreasing for
discrete $L_{\psi_{\theta}}^{-1}$.

In a case of the 
Gumbel and the Clayton copula we have \cite{hofert2008sampling}
\begin{equation}\label{eq::thitransform}
\psi_{\theta}(s) = \int_{-\infty}^{\infty} e^{-s v} d L^{-1}_{\psi_{\theta}}(v) 
= \mathbb E\left( e^{-s 
	\mathcal{V}} \right)
\end{equation}
for $s \in [0, \infty)$, where $\mathcal{V}$ is a random variable that is
distributed according to $L_{\psi_{\theta}}^{-1}$. In this case, if we extract 
most of 
information about the
correlation between $k$ changed and $n-k$ unchanged marginals into vector
$X^{\text{core}}_{k+1}$, we can carry this information (in a sense of
ordering) through Alg.~\ref{alg:arch_sampler}.  In a case of the AMH copula we
have discrete $L_{\psi_{\theta}}^{-1}$ fulfilling a discrete version of the
transform. In this case  some information may be lost, since the function in not
strictly increasing. Finally let us take a particular case of the Gumbel copula 
where 
$L^{-1}_{\psi_{\theta}}$
is a L\'evy stable distribution~\cite{hofert2008sampling} without analytical
form. Still, its element-wise sampling is discussed in \cite{mcneil2008sampling,
	nolan2003stable}. Hence we sample $t$ realisations of appropriate L\'evy 
	stable
distribution using \cite{mcneil2008sampling, nolan2003stable} and sort an
outcome according to $X^{\rm core}_{k+1}$. Such generated data is used in
Alg~\ref{alg:arch_sampler}. Since for large $t$ outcome would converge to those
from quantile function, we found our approach well motivated.
This approach can be used for other copulas not considered in this paper, for
which $L_{\psi_{\theta}}^{-1}$ is not known.

Following this discussion, our approach to preserve a correlation between
$k$ changed marginals and $n-k$ unchanged marginals concerns preparing $\mathbf
X^{\rm core}$ from data included in $k$ marginals that are changed. For the sake
of Alg.~\ref{alg:arch_sampler}, $\mathbf X^{\rm core}$ must contain $t$
realisations of $k+1$ independent uniform marginals~\cite{marshall1988families,
	hofert2008sampling}. However those marginals can be correlated with reminding
$n-k$ marginals of unchanged data, which should be preserve in changed data.
Hence in Alg.~\ref{alg::core} the eigenvector decomposition is performed in 
such a way that
$(k+1)$\textsuperscript{th} eigenvalue is highest,
\ie~$(k+1)$\textsuperscript{th} marginal carries largest information. We
include this information in $X^{\rm core}_{k+1}$. Note that 
function~\eqref{eq::fv} is strictly increasing in $x_i$ for constant $v \in 
(0,1)$ as well. Hence substituting elements of 
$X_{1}^{\text{core}},\ldots,X_{k}^{\text{core}}$ for $\mathbf{x}$ in 
Alg.~\ref{alg:arch_sampler} will carry an information through the Algorithm
as well. We call it local information since $X_{i}^{\text{core}}$ correspond to 
the $i$\textsuperscript{th} marginal of the Archimedean copula. It is in 
contrary 
to $X_{k+1}^{\text{core}}$ that carry global information via $v$ in 
Equation~\eqref{eq::fv}. Concluding, in our approach the global information is 
more significant that the local one.

\begin{algorithm}[t]
	\caption{Change a part of multivariate normal distributed data into 
	Archimedean copula 
	distributed one with similar covariance; $\mathbf 
	X[:,\texttt{ind}]$ is a shortage for `all elements with second index in 
	$\texttt{ind}$'.}
	\label{alg::malf_gen}
	\begin{algorithmic}[1]	
		\State \textbf{Input}: $\mathbf X\in\R^{t,n}$ -- $t$ realization of 
		$n$-variate normal distribution, \texttt{dist} -- label denoting 
		replacing Archimedean copula, \texttt{ind} -- variables indexes which 
		are replaced.
		\State \textbf{Output:} malformed data $\mathbf X^{\rm out}$ within 
		copula $\texttt{dist}$ on marginal subset $\texttt{ind}$. 
		\Function{arch\_copula\_malformation}{$\mathbf X$, $\texttt{dist}$, 
		$\texttt{ind}$}
		\State $\mathbf S \leftarrow$ variance diagonal matrix of $\mathbf X$
		\State $\mu \leftarrow $ mean vector of $\mathbf X$
		\State $k \leftarrow$ length of \texttt{ind}
		\For {$i=1,\ldots,t$}
		\State $\mathbf  x^{\rm std}_i \leftarrow \mathbf S^{-\frac{1}{2}} ( \mathbf x_i - \mu)$ 
		\EndFor
		
		\State $\mathbf X^U \leftarrow \textsc{cdf}_{\mathcal N(0,1)}(\mathbf 
		X^{\rm std})$ \Comment $\X^U \in [0,1]^{t \times n}$
		\State 
		
		\State $\mathbf X^{\rm core} \leftarrow \textsc{core}(\mathbf 
		X^{{\rm 
				std}}[:,\texttt{ind}])$\Comment{or with 
		\textsc{core\_naive}(\texttt{ind})}
		\State derive generator $\psi_{\theta}$ given 
		\texttt{dist}
		\State derive parameter $\theta$ based on $\mathbf X^{\rm 
			std}[:,\texttt{ind}]$
		 \For{$i=1,\dots,t$}
		\State $\mathbf x^U_i[\texttt{ind}] \leftarrow 
		\textsc{arch sampler}(\psi_{\theta}, \mathbf  x^{\rm core}_i[1,\dots,k],x^{\rm core}_{i,k+1})$ 
		\State $\mathbf x^{\rm out}_i \leftarrow 
		\mathbf S^{\frac{1}{2}}(\textsc{quantile}_{\mathcal N(0,1)}(\mathbf x^U_i) 
		+\mu)$ 

		\EndFor
		
		\State\Return $\mathbf X^{\rm out}$
		
		\EndFunction 
	\end{algorithmic}
\end{algorithm}

\begin{algorithm}[t]
	\caption{\textsc{core} function for generating initial data form 
		multivariate uniform distribution}
	\label{alg::core}
	\begin{algorithmic}[1]	
		\State \textbf{Input}: $\mathbf X_{\rm standard}\in\R^{t,k}$ - $t$ 
		realization of $k$-variate normal distribution with standard normal 
		marginals
		\State \textbf{Output:} $ \mathbf X^{\rm core}\in\R^{t,k+1}$ -- $t$ 
		realisations of $k+1$ independent $U(0,1)$. 
		\Function{core}{$\mathbf X$}
		
		\State $\mathbf X \leftarrow \mathbf X, 
		\textsc{random}_{\mathcal{N}(0,1)}(t)$ \Comment{$\in\R^{t,k+1} $, 
			append $(k+1)^{\rm th}$ column from $\mathcal{N}(0,1)$}
		\State $\Sigma \leftarrow \textsc{cor}(\mathbf X)$ 
		\State $\lambda, U \leftarrow \textsc{eigenvals}(\Sigma), 
		\textsc{eigenvec}(\Sigma)$ 
		\Comment{$\lambda_1 < \ldots < \lambda_{k+1}$}
		\For{l=1,\dots, t}
		\State $\mathbf{x}_l\leftarrow \mathbf{x}_l\cdot U$
		\For{i=1,\dots, k+1}
		\State $x_{l,i}^{\rm core} \leftarrow 
		\textsc{cdf}_{\mathcal{N}(0,\sqrt{\lambda_i})}(x_{l,i})$
		\EndFor 
		\EndFor 
		\State\Return $\X^{\rm core}$		
		\EndFunction 
	\end{algorithmic}	
\end{algorithm}

Algorithms presented here are implemented in a \texttt{Julia} programming 
language \cite{bezanson2014julia} and can be found on a GitHub repository 
\cite{cop}, see \texttt{gcop2arch} function therein.

\subsection{MEV vs JSBS algorithms}\label{s::mevjsbs}

In this section we show, that our algorithm generates data which distinguishes
MEV and JSBS algorithms. Our experiment is based on $100$-dimensional vector. We
chose random covariance matrix $\Sigma$ with ones on a diagonal, and generate
$10^5$ multivariate normal samples with $0$ mean vector and covariance $\Sigma$.
The covariance matrix was chosen as follows. First we chose $\mathbf{M}\in 
\R^{n\times
	n}$, which elements are distributed according to uniform distribution $U(0,1)$.
Let $\mathbf{D_M}=\operatorname{diagm}(\mathbf{M} \mathbf{M}^{\top})$ be 
diagonal matrix of $\mathbf{M} \mathbf{M}^\top$. Our
covariance matrix take form $\Sigma=\mathbf{D_{M}}^{-\frac{1}{2}}\mathbf{M} 
\mathbf{M}^\top
\mathbf{D_{M}}^{-\frac{1}{2}}$.

Then 8 random variables were changed according to our algorithm, using both 
\textsc{core} presented in Alg.~\ref{alg::core} and  
$\textsc{core\_naive}$, which generates samples according to uniform 
distribution. First kind of samples was considered using MEV and JSBS algorithms, 
second kind were analysed using MEV algorithm only. 

\begin{figure}[h!]
	\centering		
\subfloat[Number of samples, for which given numbered of malformed variables were found\label{subfig:itworks}]
{\includegraphics[]{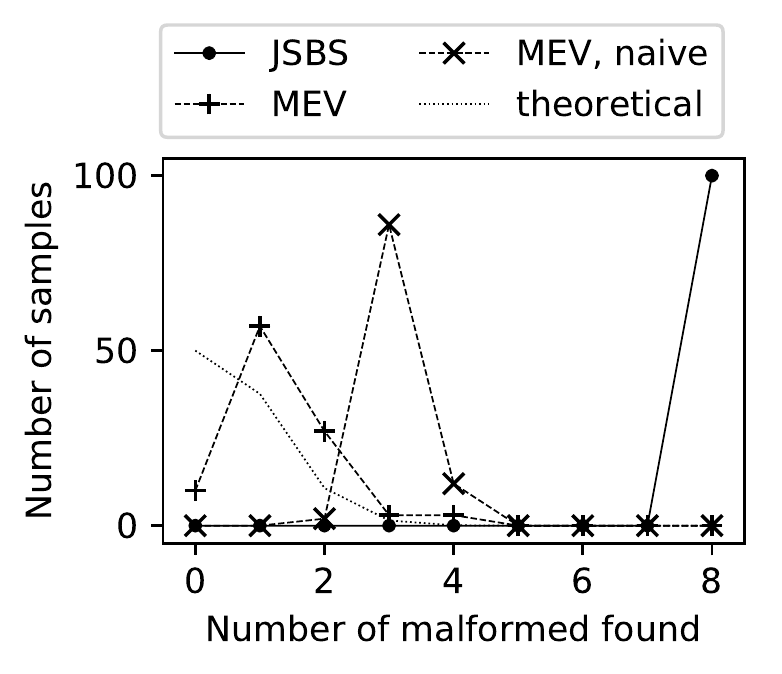}}
\subfloat[Relative covariance change for all samples\label{subfig:itworks-covariances}]{
\includegraphics[]{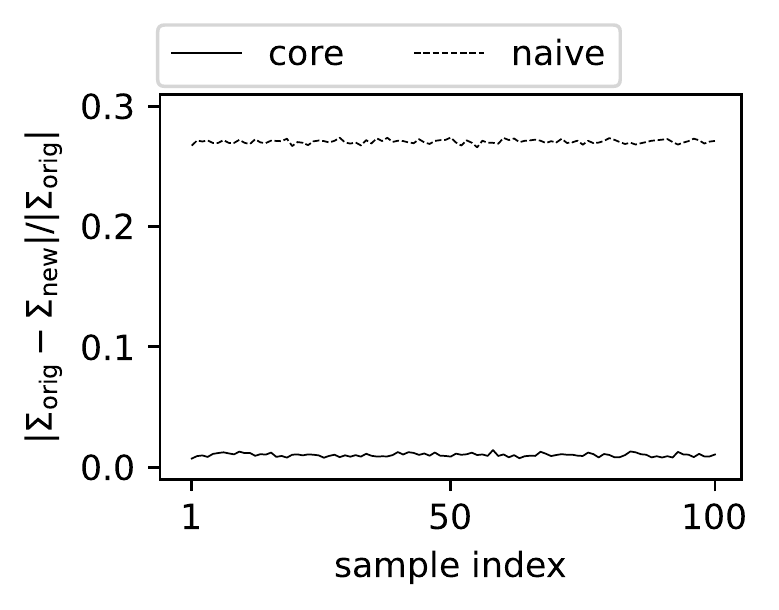}}
\caption{Comparison of the core and the naive algorithms on data malformed 
using Gumbel 
copula. 
\ref{subfig:itworks}: number of experiments, for which given number of 
malformed marginals were found. `theoretical' is the theoretical number of 
`detected' marginals in the case of a random guess. In all cases we 
have malformed $8$ marginals an analyse $8$ detected ones (left from the 
elimination procedure). 
\ref{subfig:itworks-covariances}: relative change of covariance matrix 
calculated 
according to $\|\Sigma_{\rm orig} - \Sigma_{\rm new}\| / \|\Sigma_{\rm orig}\|$ 
for all experiments.}\label{fig:it-works}
\end{figure}

In Fig.~\ref{fig:it-works} we present exemplary results on data malformed using
Gumbel copula. The figure presents how many from 8 detected marginals are the
malformed one. We observe, that in each case the JSBS algorithm have detected
all malformed marginals. The MEV yields result similar to the random guess
algorithm. Contrary we can see, that malformed data  by means of the naive
algorithm \textsc{core\_naive} were partially detected by the MEV. We explain
this fact by the large influence of the naive algorithm on the covariance
matrix, see Fig.~\ref{subfig:itworks-covariances}. As discussed in the previous
subsection, in the naive algorithm case we have zero correlation between
malformed and not malformed subsets, which is not true in the \texttt{core}
algorithm case.

\subsection{Algorithm analysis}

\begin{figure}[h!]
	\centering		
	\subfloat[\label{subfig:mev-notnested}efficiency of MEV algorithm on malformed data using different copulas]
	{\includegraphics[]{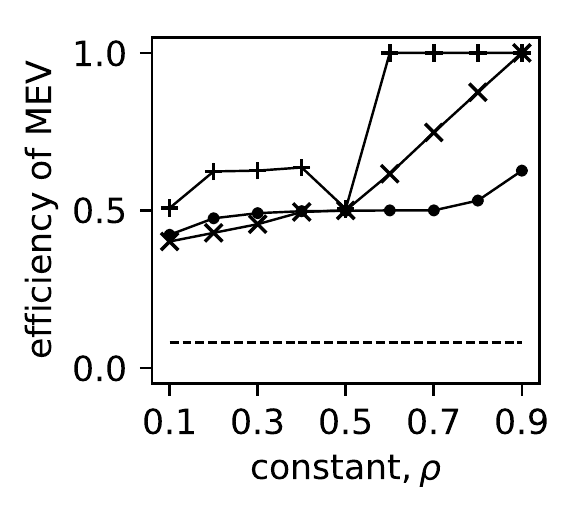}
		\includegraphics[]{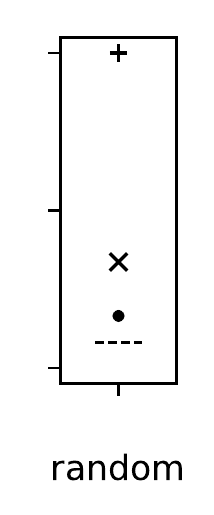}
		\includegraphics[]{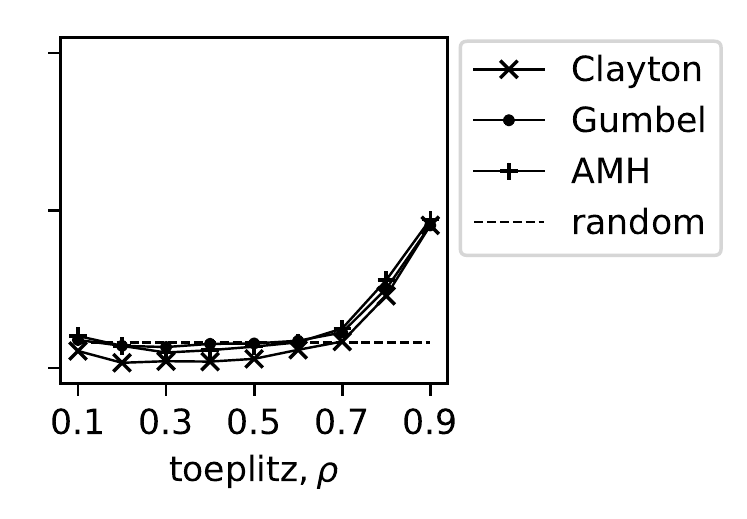}}\\
	\subfloat[\label{subfig:jsbs-notnested}efficiency of JSBS algorithm on malformed data using different copulas]
	{\includegraphics[]{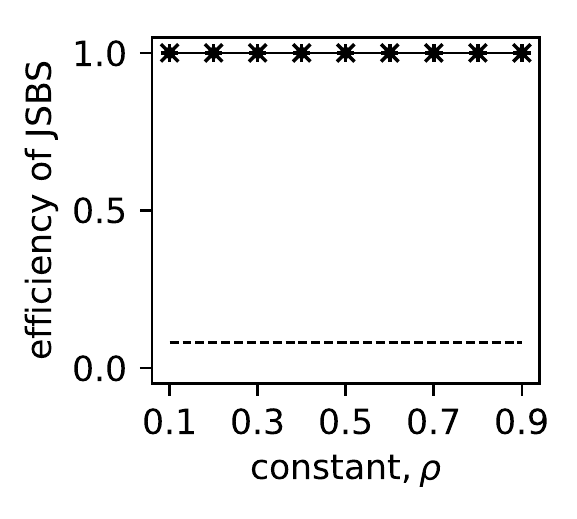}
		\includegraphics[]{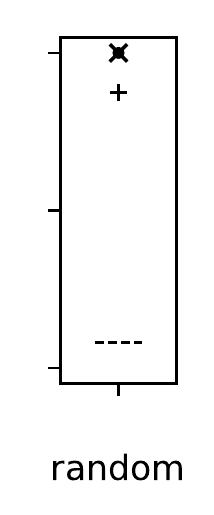}
		\includegraphics[]{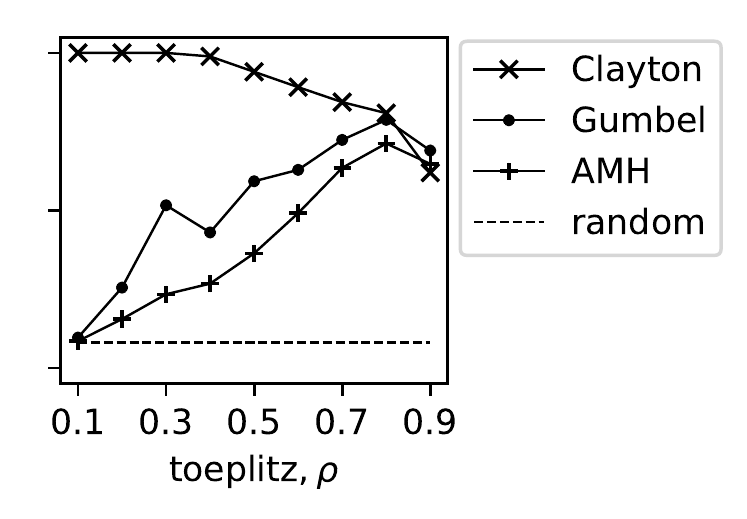}}
	
	\caption{Comparison of \ref{subfig:mev-notnested} MEV and  
	\ref{subfig:jsbs-notnested} JSBS algorithms on malformed data using the 
	core algorithm, different copulas, and different methods of covariance 
	matrices generation. Note 
	for constant and random covariance matrices JSBS achieve full efficiency. } 
	\label{fig:pure-cor-matrix}
\end{figure}
\begin{figure}[h!]
	\centering		
	\subfloat[\label{subfig:mev-noised}MEV 
	algorithm]{\includegraphics[]{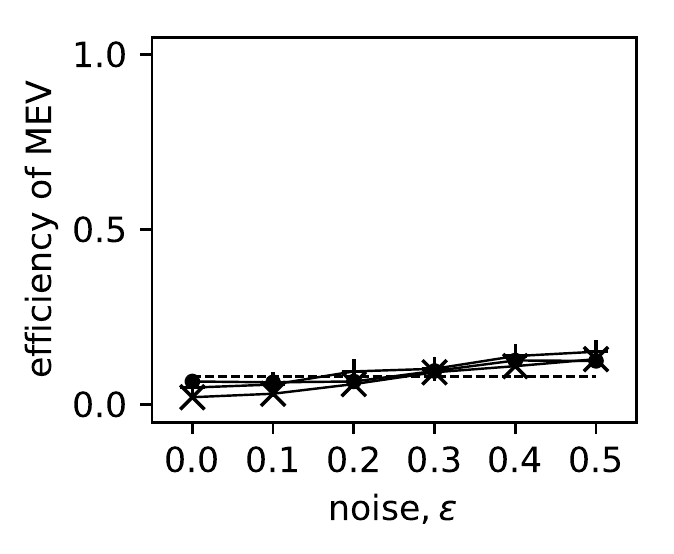}}
	\subfloat[\label{subfig:jsbs-noised}JSBS 
	algorithm]{\includegraphics[]{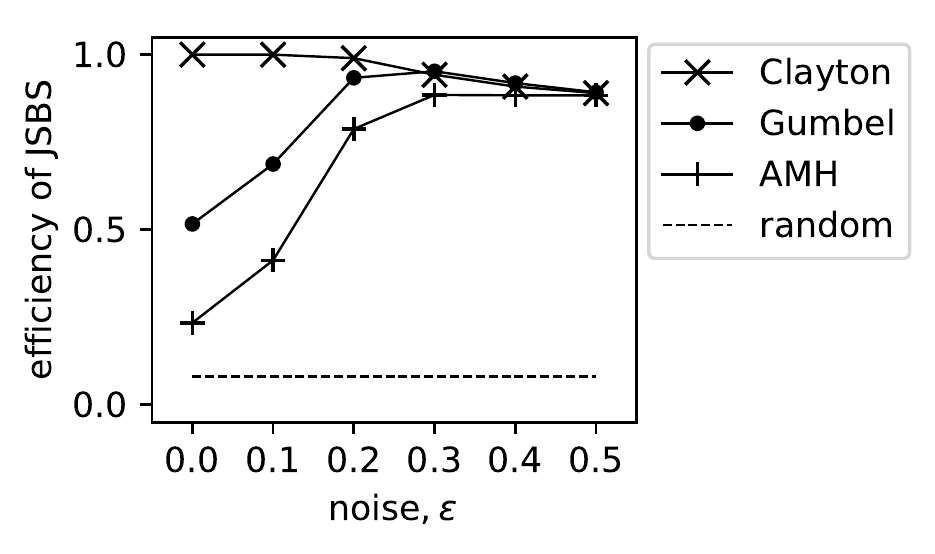}}
	\caption{Efficiency of \ref{subfig:mev-noised} the MEV and 
	\ref{subfig:jsbs-noised} the JSBS algorithm for noised Toeplitz correlation 
	matrix with parameter $\rho =0.3$. Note that the MEV discriminability is 
	not 
	affected significantly by the noise parameter $\varepsilon$ value, while 
	the JSBS discriminability in general rises with the noise 
	strength.}\label{fig:noise} 
\end{figure}

In this section we analyse our algorithms using Clayton, Gumbel and AMH
Archimedean copulas and different covariance matrices generators. In particular we have analysed 
\begin{enumerate}
	\item true random covariance matrix  $\Sigma^{\rm r}$, generated as 
	described in Section~\ref{s::mevjsbs},
	\item constant correlation matrix $\Sigma^{\rm c}_\alpha$, where the 
	correlation between two different marginal random variables equal to a free 
	parameter $\alpha \in [0,1]$,
	\item Toeplitz correlation matrix $\Sigma^{\rm T}_\rho$, where the 
	correlation between $i$-th and $j$-th elements equals $\rho^{|i-j|}$ for 
	$\rho \in [0,1]$,
	\item noised version of Toeplitz correlation matrix $\Sigma^{\rm T }_{\rho, \varepsilon}$ \cite{hardin2013method}, where $\varepsilon$ corresponds to noise impact.
\end{enumerate}
Our measure of goodness of
discrimination is the percentage of correctly discriminated variables, \ie{} if
$p^{\rm th}$ experiment has correctly recognized $r_p$ malformed variables, 
then the
mean discriminability is defined as
\begin{equation}
\bar D = \sum_{p=1}^{100} \frac{r_p}{8}.
\end{equation}
Note that $\bar D=1$ is achieved when all malformed variables were correctly
recognized for all experiments. Remark, that random guess would give the
expectedly discriminability $\bar{D}_{\rm theor}= 0.08$. We claim, that our
algorithm works well if the discriminability of MEV algorithm is close to the
theoretical, while for JSBS it is much higher.

Let us analyze the algorithm. In Fig.~\ref{fig:pure-cor-matrix} we show that its
efficiency depends both on the choice of the correlation matrix generation
method and the Archimedean copula used in the algorithm. For constant
correlation matrix, the JSBS achieves full discriminability for all copulas,
however the MEV still produces discriminability much higher than the random
choice, especially in the AMH copula case given the $\rho > 0.5$ parameter of a
constant correlation matrix. For random correlation matrix generation, JSBS
achieved almost full discriminability for all copulas. In the case of MEV the
efficiency was high for AMH, but small for Gumbel copula. 

Note, that the AMH copula can achieve only $\rho \in (0,0.5)$ correlation, hence
covariance matrices, implying high overall correlations, may be affected by the
algorithm with the AMH copula. Further as discussed in
subsection~\ref{sub::datamalf} the algorithm with the AMH copula may lose some
information due to the discrete inverse Laplace–Stieltjes transform of the AMH
copula generator. This observation may explain higher detectability by the MEV
of data malformed using the AMH copula and constant correlation matrix
parametrised by $\rho < 0.5$.

The best stability of results can be observed for Toepltiz correlation matrices.
Here the MEV achieves similar efficiency as the random choice for $\rho \leq
0.7$. The JSBS achieves good discriminability, for all copulas, starting from
the Toeplitz parameter $\rho = 0.3$. Here the Clayton copula produces almost
full efficiency.

In Fig.~\ref{fig:noise} we present impact of noise in the Toeplitz correlation
matrix, on the detection performance. We observe that MEV performance does not
depend on the noise parameter significantly. This is not the case for the JSBS,
where in the case of Gumbel and AMH copulas the larger the noise parameter
$\varepsilon$, the higher the discriminability is. The Clayton copula outcomes are
good for all values of the $\varepsilon$.

To conclude, the introduction of noise improves JSBS's and MEV's detection, but
in the latter the case the change is negligible. In our opinion, this is due to
the fact, that we lose some information while performing data malformed using
the core algorithm, however this lose is small enough to be hidden in the noise.

\section{Conclusions} 

\label{sec:conclusions} In this paper, we presented and analysed algorithm, 
which replaces a chosen subset of marginals, initially distributed by a 
multivariate Gaussian distribution by those distributed according to an 
Archimedean copula. While our algorithm is designed for particular Archimedean 
families, it can be easily generalised to other Archimedean copulas. The 
correctness of our algorithm was numerically confirmed by comparing two 
distinct feature selection algorithms: the JSBS that utilises higher order 
correlations and the MEV that does not utilise higher order correlations. The 
comparison was performed for a various choice of correlation matrices. While 
for some correlation matrices MEV has provided almost random results, JSBS in 
most cases has given almost full discrimination. Hence in our opinion, the 
algorithm can be used to provide data for analysis of various complicated 
machine-learning algorithms. Such analysis would determine, whether the 
algorithm utilises higher order correlations or not. One should note that our 
algorithm does not affect a covariance matrix significantly and leaves all 
univariate marginal distributions unchanged as well as it does not reshuffle 
realisations of data. It is why the algorithm does not introduce other factors 
apart from higher order correlations.

The resulting algorithm can be generalised in various direction. First, we 
believe that the effect on a covariance matrix can be diminished, this would 
reduce discriminability of the MEV algorithm. Second, the algorithm can be 
generalised into a more extensive collection of copulas, not necessarily 
involving Archimedean ones. We can mention here the Marshall-Olkin bivariate 
copula or the Fr\'echet maximal copula derived from the Fr\'echet–Hoeffding 
upper copula bound.

\paragraph{Acknowledgments}
The research was partially financed by the National Science Centre, Poland --
project number 2014/15/B/ST6/05204. The authors would like to thank Jarosław 
Adam Miszczak for revising the manuscript and discussion.

\bibliographystyle{ieeetr}
\bibliography{copulas}

\begin{thebibliography}{10}

\bibitem{fan2011sparse}
J.~Fan, J.~Lv, and L.~Qi, ``Sparse high-dimensional models in economics,'' {\em
  Annual Review of Economics}, vol.~3, pp.~291--317, 2011.

\bibitem{ando2017clustering}
T.~Ando and J.~Bai, ``Clustering huge number of financial time series: A panel
  data approach with high-dimensional predictors and factor structures,'' {\em
  Journal of the American Statistical Association}, vol.~112, no.~519,
  pp.~1182--1198, 2017.

\bibitem{dai2015feature}
Y.~Dai, B.~Hu, Y.~Su, C.~Mao, J.~Chen, X.~Zhang, P.~Moore, L.~Xu, and H.~Cai,
  ``{Feature selection of high-dimensional biomedical data using improved SFLA
  for disease diagnosis},'' in {\em Bioinformatics and Biomedicine (BIBM), 2015
  IEEE International Conference on}, pp.~458--463, IEEE, 2015.

\bibitem{li2017clustering}
W.~Li, G.~Wang, and K.~Li, ``{Clustering algorithm for audio signals based on
  the sequential Psim matrix and Tabu Search},'' {\em EURASIP Journal on Audio,
  Speech, and Music Processing}, vol.~2017, no.~1, p.~26, 2017.

\bibitem{cordes2015feature}
K.~Cordes, L.~Grundmann, and J.~Ostermann, ``Feature evaluation with
  high-resolution images,'' in {\em International Conference on Computer
  Analysis of Images and Patterns}, pp.~374--386, Springer, 2015.

\bibitem{roweis2000nonlinear}
S.~T. Roweis and L.~K. Saul, ``Nonlinear dimensionality reduction by locally
  linear embedding,'' {\em science}, vol.~290, no.~5500, pp.~2323--2326, 2000.

\bibitem{duda2012pattern}
R.~O. Duda, P.~E. Hart, and D.~G. Stork, {\em Pattern classification}.
\newblock John Wiley \& Sons, 2012.

\bibitem{jolliffe2002principal}
I.~T. Jolliffe, ``Principal components as a small number of interpretable
  variables: some examples,'' {\em Principal Component Analysis}, pp.~63--77,
  2002.

\bibitem{jondeau2017moment}
E.~Jondeau, E.~Jurczenko, and M.~Rockinger, ``{Moment component analysis: An
  illustration with international stock markets},'' {\em {Journal of Business
  \& Economic Statistics}}, pp.~1--23, 2017.

\bibitem{arismendi2014monte}
J.~C. Arismendi and H.~Kimura, ``{Monte Carlo Approximate Tensor Moment
  Simulations},'' {\em {Available at SSRN 2491639}}, 2014.

\bibitem{becker2014eeg}
H.~Becker, L.~Albera, P.~Comon, M.~Haardt, G.~Birot, F.~Wendling, M.~Gavaret,
  C.-G. B{\'e}nar, and I.~Merlet, ``{EEG extended source localization:
  tensor-based vs. conventional methods},'' {\em NeuroImage}, vol.~96,
  pp.~143--157, 2014.

\bibitem{geng2011research}
M.~Geng, H.~Liang, and J.~Wang, ``{Research on methods of higher-order
  statistics for phase difference detection and frequency estimation},'' in
  {\em Image and Signal Processing (CISP), 2011 4th International Congress on},
  vol.~4, pp.~2189--2193, IEEE, 2011.

\bibitem{scholkopf1998nonlinear}
B.~Sch{\"o}lkopf, A.~Smola, and K.-R. M{\"u}ller, ``Nonlinear component
  analysis as a kernel eigenvalue problem,'' {\em Neural computation}, vol.~10,
  no.~5, pp.~1299--1319, 1998.

\bibitem{hoffmann2007kernel}
H.~Hoffmann, ``{Kernel PCA for novelty detection},'' {\em Pattern recognition},
  vol.~40, no.~3, pp.~863--874, 2007.

\bibitem{baudat2000generalized}
G.~Baudat and F.~Anouar, ``Generalized discriminant analysis using a kernel
  approach,'' {\em Neural computation}, vol.~12, no.~10, pp.~2385--2404, 2000.

\bibitem{lee2007nonlinear}
J.~A. Lee and M.~Verleysen, {\em Nonlinear dimensionality reduction}.
\newblock Springer Science \& Business Media, 2007.

\bibitem{sakurada2014anomaly}
M.~Sakurada and T.~Yairi, ``Anomaly detection using autoencoders with nonlinear
  dimensionality reduction,'' in {\em Proceedings of the MLSDA 2014 2nd
  Workshop on Machine Learning for Sensory Data Analysis}, p.~4, ACM, 2014.

\bibitem{wang2014generalized}
W.~Wang, Y.~Huang, Y.~Wang, and L.~Wang, ``Generalized autoencoder: A neural
  network framework for dimensionality reduction,'' in {\em Proceedings of the
  IEEE conference on computer vision and pattern recognition workshops},
  pp.~490--497, 2014.

\bibitem{hinton2006reducing}
G.~E. Hinton and R.~R. Salakhutdinov, ``Reducing the dimensionality of data
  with neural networks,'' {\em science}, vol.~313, no.~5786, pp.~504--507,
  2006.

\bibitem{mcneil2009multivariate}
A.~J. McNeil and J.~Ne{\v{s}}lehov{\'a}, ``Multivariate archimedean copulas,
  d-monotone functions and $l_1$-norm symmetric distributions,'' {\em The
  Annals of Statistics}, pp.~3059--3097, 2009.

\bibitem{sheffield1985selecting}
C.~Sheffield, ``Selecting band combinations from multispectral data,'' {\em
  Photogrammetric Engineering and Remote Sensing}, vol.~51, pp.~681--687, 1985.

\bibitem{geng2015joint}
X.~Geng, K.~Sun, L.~Ji, H.~Tang, and Y.~Zhao, ``{Joint Skewness and Its
  Application in Unsupervised Band Selection for Small Target Detection},''
  {\em Scientific reports}, vol.~5, 2015.

\bibitem{bezanson2014julia}
J.~Bezanson, A.~Edelman, S.~Karpinski, and V.~B. Shah, ``{Julia: A fresh
  approach to numerical computing},'' {\em SIAM Review}, vol.~59, no.~1,
  pp.~65--98, 2017.

\bibitem{cop}
K.~Domino and A.~Glos, ``{DatagenCopulaBased.jl}.''
  https://doi.org/10.5281/zenodo.1213710, 2018.

\bibitem{mccullagh2009cumulants}
P.~McCullagh and J.~Kolassa, ``{Cumulants},'' {\em Scholarpedia}, vol.~4,
  no.~3, p.~4699, 2009.

\bibitem{schatz2014exploiting}
M.~D. Schatz, T.~M. Low, R.~A. van~de Geijn, and T.~G. Kolda, ``{Exploiting
  symmetry in tensors for high performance: Multiplication with symmetric
  tensors},'' {\em SIAM Journal on Scientific Computing}, vol.~36, no.~5,
  pp.~C453--C479, 2014.

\bibitem{domino2017tensorsnet}
K.~Domino, {\L}.~Pawela, and P.~Gawron, ``Efficient computation of higer-order
  cumulant tensors,'' {\em SIAM Journal on Scientific Computing}, vol.~40,
  no.~3, pp.~A1590--A1610, 2018.

\bibitem{kendall1946advanced}
M.~G. Kendall {\em et~al.}, ``{The advanced theory of statistics},'' {\em The
  advanced theory of statistics.}, no.~2nd Ed, 1946.

\bibitem{lukacs1970characteristics}
E.~Lukacs, ``Characteristics functions,'' {\em Griffin, London}, 1970.

\bibitem{nelsen1999introduction}
R.~B. Nelsen, {\em An introduction to copulas}.
\newblock Springer Science \& Business Media, 2007.

\bibitem{cherubini2004copula}
U.~Cherubini, E.~Luciano, and W.~Vecchiato, {\em Copula methods in finance}.
\newblock John Wiley \& Sons, 2004.

\bibitem{embrechts2001modelling}
P.~Embrechts, F.~Lindskog, and A.~McNeil, ``Modelling dependence with
  copulas,'' {\em Rapport technique, D{\'e}partement de math{\'e}matiques,
  Institut F{\'e}d{\'e}ral de Technologie de Zurich, Zurich}, 2001.

\bibitem{naifar2011modelling}
N.~Naifar, ``{Modelling dependence structure with Archimedean copulas and
  applications to the iTraxx CDS index},'' {\em Journal of Computational and
  Applied Mathematics}, vol.~235, no.~8, pp.~2459--2466, 2011.

\bibitem{domino2014use}
K.~Domino and T.~B{\l}achowicz, ``The use of copula functions for modeling the
  risk of investment in shares traded on the {W}arsaw {S}tock {E}xchange,''
  {\em Physica A: Statistical Mechanics and its Applications}, vol.~413,
  pp.~77--85, 2014.

\bibitem{zhang2012application}
Q.~Zhang, J.~Li, and V.~P. Singh, ``{Application of Archimedean copulas in the
  analysis of the precipitation extremes: effects of precipitation changes},''
  {\em Theoretical and applied climatology}, vol.~107, no.~1-2, pp.~255--264,
  2012.

\bibitem{tsakiris2015flood}
G.~Tsakiris, N.~Kordalis, and V.~Tsakiris, ``Flood double frequency analysis:
  2d-archimedean copulas vs bivariate probability distributions,'' {\em
  Environmental Processes}, vol.~2, no.~4, pp.~705--716, 2015.

\bibitem{zeng2014copulas}
X.~Zeng, J.~Ren, Z.~Wang, S.~Marshall, and T.~Durrani, ``Copulas for
  statistical signal processing (part i): Extensions and generalization,'' {\em
  Signal Processing}, vol.~94, pp.~691--702, 2014.

\bibitem{peters2014communications}
G.~W. Peters, T.~A. Myrvoll, T.~Matsui, I.~Nevat, and F.~Septier,
  ``Communications meets copula modeling: Non-standard dependence features in
  wireless fading channels,'' in {\em Signal and Information Processing
  (GlobalSIP), 2014 IEEE Global Conference on}, pp.~1224--1228, IEEE, 2014.

\bibitem{silva2014statistically}
R.~F. Silva, S.~M. Plis, T.~Adal{\i}, and V.~D. Calhoun, ``A statistically
  motivated framework for simulation of stochastic data fusion models applied
  to multimodal neuroimaging,'' {\em NeuroImage}, vol.~102, pp.~92--117, 2014.

\bibitem{kumar2010probability}
P.~Kumar, ``Probability distributions and estimation of {A}li-{M}ikhail-{H}aq
  copula,'' {\em Applied Mathematical Sciences}, vol.~4, no.~14, pp.~657--666,
  2010.

\bibitem{schweizer1981nonparametric}
B.~Schweizer and E.~F. Wolff, ``On nonparametric measures of dependence for
  random variables,'' {\em The annals of statistics}, pp.~879--885, 1981.

\bibitem{de2012moments}
E.~de~Amo, M.~D. Carrillo, J.~F. S{\'a}nchez, and A.~Salmer{\'o}n, ``Moments
  and associated measures of copulas with fractal support,'' {\em Applied
  Mathematics and Computation}, vol.~218, no.~17, pp.~8634--8644, 2012.

\bibitem{marshall1988families}
A.~W. Marshall and I.~Olkin, ``Families of multivariate distributions,'' {\em
  Journal of the American statistical association}, vol.~83, no.~403,
  pp.~834--841, 1988.

\bibitem{hofert2008sampling}
M.~Hofert, ``Sampling {A}rchimedean copulas,'' {\em Computational Statistics \&
  Data Analysis}, vol.~52, no.~12, pp.~5163--5174, 2008.

\bibitem{mcneil2008sampling}
A.~J. McNeil, ``Sampling nested {A}rchimedean copulas,'' {\em Journal of
  Statistical Computation and Simulation}, vol.~78, no.~6, pp.~567--581, 2008.

\bibitem{nolan2003stable}
J.~Nolan, {\em Stable distributions: models for heavy-tailed data}.
\newblock Birkhauser New York, 2003.

\bibitem{hardin2013method}
J.~Hardin, S.~R. Garcia, and D.~Golan, ``A method for generating realistic
  correlation matrices,'' {\em The Annals of Applied Statistics},
  pp.~1733--1762, 2013.

\end{thebibliography}

\end{document}